\documentclass[preprint,12pt]{elsarticle}

\usepackage{amssymb}

\journal{Applied mathematics and computations}

\begin{document}

\begin{frontmatter}



\title{Separative power of an optimised concurrent gas centrifuge.}


\author{S.V. Bogovalov, V.D. Borman }
\ead{svbogovalov@mephi.ru}
\address{National research nuclear university (MEPHI), Kashirskoje shosse, 31, 115409, Moscow, Russia}

\begin{abstract}
The problem of  separation of uranium isotopes in a concurrent gas centrifuge is solved analytically. 
Separative power of the  optimized concurrent gas centrifuges equals to $\delta U=12.7(V/700~{\rm m/s})^2 (300 ~{\rm K}/T)L, ~{\rm kg ~SWU/yr}$, where $L$ and $V$ are  the length and linear velocity of the rotor of the gas centrifuge, $T$ is the temperature.   
This formula  well agrees with an empirical separative power of counter current gas centrifuges. The optimal value of the separative power is not unique 
on the plane $(p_w,v_z)$, where $p_w$ is 
pressure at the wall of the rotor  and $v_z$ is axial velocity of the gas.  This value is constant  on a line defined by the equation $p_wv_z=constant$. 
Equations defining the mass flux and the electric power necessary to support the rotation of the gas centrifuge are obtained.
\end{abstract}

\begin{keyword}
gas centrifuge \sep isotope separation \sep diffusion in strong centrifugal field \sep separative power 
\PACS 28.60.+s \sep 47.32.Ef\sep51.10+y\sep51.20.+d


\end{keyword}

\end{frontmatter}


\section{Introduction}
\label{s1}
{\bf Separation of isotopes has traditionally been carried out using gaseous centrifugation.  Only recently the problem of sedimentation of isotopes in liquids and solids due to diffusion of the isotopes in strong centrifugal fields attracts attention of specialists in connection with design  of the centrifuges which can provide the centrifugal field $\sim 10^6~\rm g$ at the temperature achieving $500 ^o C$ \cite{mash0}. The diffusion process in the condensed matter differs essentially from the diffusion in the gas. 
Theoretical grounds of diffusion in condensed media are still under debates (see \cite{br2} and reference therein). Specific features of the sedimentation of liquid and solid mixtures in strong centrifugal fields and corresponding experiments were discussed in \cite{mash1,mash2,mash3}. The attempts to interpret  these experiments  can be found in \cite{wirzba1} and references therein. In the present paper we discuss one of the unsolved key problem  of the traditional process of separation of uranium isotopes in the gas centrifuges (hereafter GC). 

The transport of gaseous isotopes in centrifugal fields of the order of $10^6 ~\rm g$  is rather specific in compare with the transport of the isotopes in a condensed media. In so strong fields the density of the gas varies on $\sim 10^{5}- 10^6$ orders of magnitude per centimetre along radius. Convection in the GC plays crucial role because it is artificially excited to increase the separative power of the GC \cite{borman}. The relationship between the diffusive flux and the flux due to convection also varies on the same 
$\sim 10^{5}- 10^6$ orders of magnitude. This complicates the solution of the problem because it is necessary to solve all the system of Navier-Stokes equations and plus to solve the diffusion equation for the mixture of isotopes in extreme conditions of strong variation of density of the gas. On this reason the problem of the isotope separation is solved up to now only numerically.}

Separation of heavy isotopes in gas centrifuges is used for industrial production of enriched uranium starting with the middle of past century. 
In spite of old history, the problem of separative power of an optimized GC is not solved yet. Solution of this problem is interesting from practical point of view.   Knowledge or estimate of the separative power of GC is important for development of new more efficient GC and important for experts  dealing with the problem of proliferation prevention of the separation technology.

Attempts to estimate the separative power of GC has been made firstly by Dirac  \cite{cohen}. He has shown that the separative power $\delta U_{max}$ of any GC can not exceed the value
\begin{equation}
 \delta U_{max}={\pi \rho DL\over 2} ({\Delta M V^2\over 2RT})^2,   
\label{eq1}
\end{equation}
where $\rho D$ is the density of uranium hexafluoride ($UF_6$) times the coefficient of self-diffusion for uranium isotopes $^{238}U$ and $^{235}U$. $\Delta M$ is the mass difference between two uranium isotopes , $R$ is the gas-law constant, T is gas temperature, L is the length of the rotor of GC, $V$ is  linear velocity of the rotor rotation.

At the beginning of 1960's an Onsager group from US developed a theory called  ‘‘the pancake approximation’’ that reduced the problem of the GC gas dynamics to solution of linear elliptical equations of the sixth-order partial derivatives for two variables  derivatives \cite{onsager,du}.  This approach gave the following equation for the separative power of the GC
\begin{equation}
 \delta U= (0.038 V-11.5)L, ~kg\cdot SWU/yr.
\label{eq2}
\end{equation}
It is important to note that in contrast to eq. (\ref{eq1}) where the separative power increases as  $V^4$, in eq.( \ref{eq2}) the separative power growths linearly with $V$. 

Experimental data collected with help of large amount of Russian GC shine new light on this question. According to \cite{senchenko} the separative power is defined by the following empirical equation
\begin{equation}
 \delta U=12L \left({V\over 700 {\rm v/s}}\right)^2\left({2a\over 12{\rm cm}}\right)^{0.4}, ~kg \cdot SWU/yr,
\label{eq3}
\end{equation}
where $L$ is measured in meters. Recently this result has been well confirmed   by more extended experimental data \cite{yatsenko}.

In the empirical eq. (\ref{eq3}) $\delta U \sim V^2$. 
This  dramatically contradicts to simple  theoretical arguments. At relatively small separation coefficient $q$ defined as a ratio of the concentration of $^{235}U$ in the product flux over the concentration in the waste flux the separative power equals to 
\begin{equation}
 \delta U=\theta(1-\theta){F(q-1)^2\over 2}, 
\label{eq4}
\end{equation}
where $\theta=P/F$ is the ratio of the product mass flux $P$ over feed mass flux $F$  \cite{borman}. Radial separation in the centrifugal field gives the following   
dependence of $q$ on $V$
\begin{equation}
 q=\exp{\Delta M \gamma V^2\over 2RT},
\end{equation}
which  unambiguously gives that $\delta U \sim V^4$.  This dependence takes place in (\ref{eq1}) but do not agree with the experiment. 
For many years this problem remains a challenge for specialists.  Recently   a new equation defining the separative power of GC has been  proposed in \cite{kemp} 
\begin{equation}
 \delta U= \left({V^2 L\over 33000}\right) e_E,~ kg\cdot SWU/yr,
\label{eq6}
\end{equation}
where $V$ is measured in $\rm m/s$, $L$ is rotor length in meters, $e_E$ - some numerical coefficient. This equation already correctly reproduces the empirical law (\ref{eq3}). Unfortunately eq. (\ref{eq6}) has been obtained exploring  weakly controlled assumptions and details of calculations are unknown. Therefore, the question about dependence of the separative power of the GC on the parameters remains opened up to now.  

It is necessary to stress that we discuss here the separative power of GC optimized on all  parameters which can be controlled by a designer. 
The separative power is the function of a lot of parameters $\delta U(V,L,T,a, \alpha_1, \alpha_2,...)$, where the series of parameters $\alpha_i$ includes for example pressure at the wall of the rotor, $F$,$\theta$, variation of temperature along the rotor  $\delta T$ and many other.  Optimization of the GC is reduced to a search of maximum of this function at the variation of all the parameters $\alpha_i$.  Such a search is performed for every series of $V,~L,~T,~a$. Therefore, the separative power of the optimized GC depends only on the  limited set of the parameters $V,~L,~T,~a$.  Such a formulation of the problem carries additional difficulties in the solution of the problem because it is necessary not only to calculate the separative power of the GC but additionally to optimize (to find maximal value) in relation to all possible parameters at fixed $V$, $L$, $a$ and $T$. 

In this paper we give the solution of the problem of the separative power of the optimized concurrent GC  from primary equations. This type of GC has been considered firstly in \cite{cohen}, where the separative power of this type of GC has been estimated as
\begin{equation}
\delta U=0.166 \cdot 2\pi \rho DL({\Delta M V^2\over 2RT})^2,
\end{equation}
which is only 66\% less the maximal possible separative power given by (\ref{eq1}) and well agree  with the  arguments above in favour $V^4$ dependence of $\delta U$  . 

{\bf In this work we give analytical equation for the separative power of an  optimized concurrent gas centrifuge for the first time. In contrast to the previous  results we show that the correct solution gives another functional dependence of the separative power on $V$ and $L$ close to the empirical one given by eq. (\ref{eq3})}. 

The paper is organized as follows. In the second section we present the scheme of the concurrent centrifuge, basic equations and assumptions.  In sec. 3 the solution is described in details. In sec. 4 the optimized separative power is calculated and finally we discuss the solution in  last sec.5.

\section{Isotope separation in the concurrent centrifuge}

\subsection{Hydrodynamics}

\begin{figure} 
\begin{center}
\includegraphics[width=110mm]{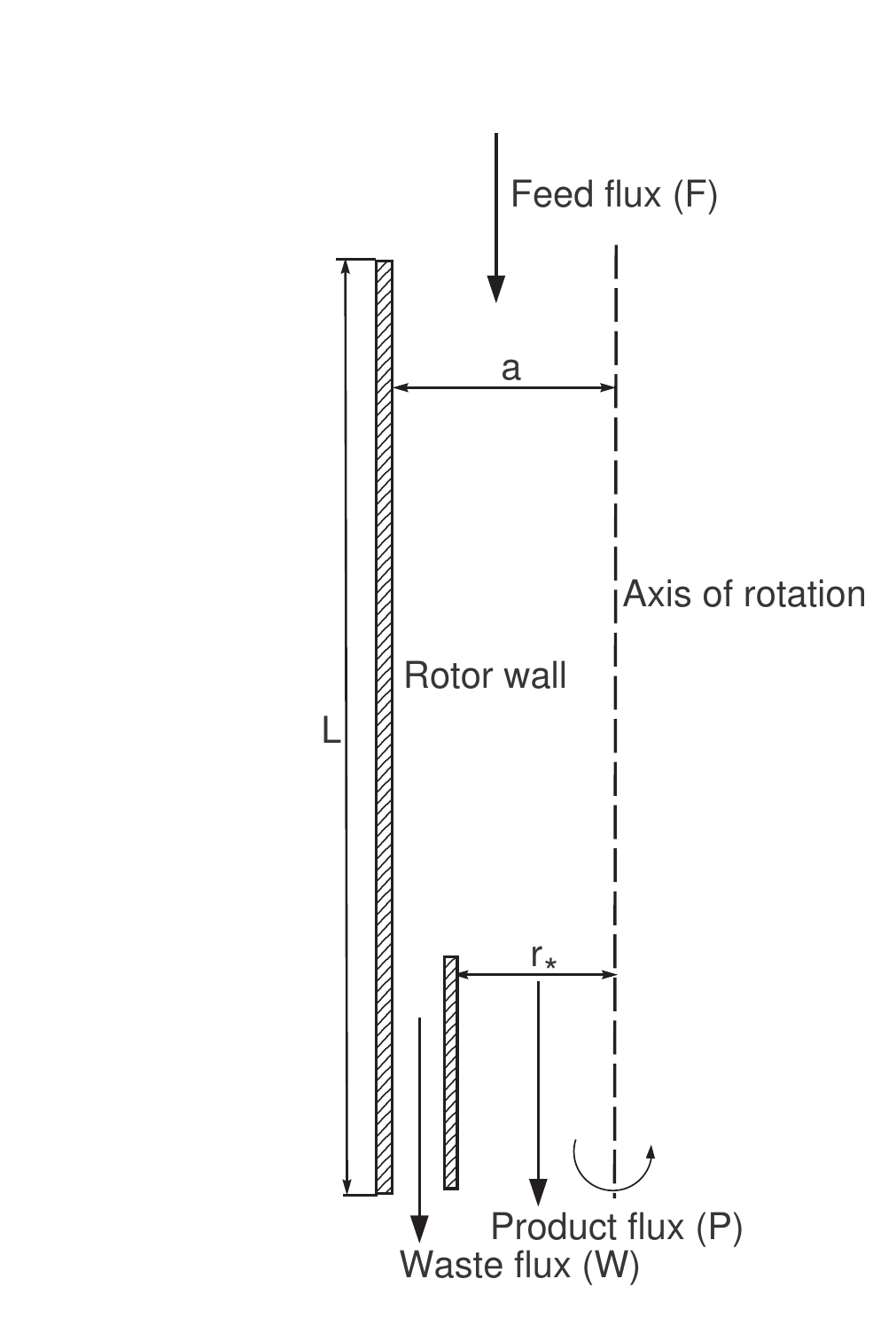}
\caption{A scheme of the concurrent centrifuge. The working gas is feed from the top of the rotor (inlet) with the distribution of pressure and velocity corresponding to the rigid body rotation of the gas. The product and waste fluxes leave the rotor at the bottom (outlet) and are separated by a  concentric cylinder with radius $r_*$. }
 \label{f1}
\end{center}
\end{figure}

The schematic view on the concurrent GC is shown in fig. \ref{f1}. Here we use a scheme intentionally simplified for the numerical analysis. The working gas $UF_6$ is load into the GC with feed flux $F$ from the top (inlet) with the velocity $v_z$ independent on the radius $r$. 
We assume that the pressure $p$ and density $\rho$  correspond to the full hydrodynamic equilibrium in the radial direction. They are 
\begin{equation}
 p=p_w\exp{\left({\gamma V^2\over 2C^2}(({r\over a})^2-1)\right)},
\label{eq8}
\end{equation}
and 
\begin{equation}
 \rho=\rho_w\exp{\left({\gamma V^2\over 2C^2}(({r\over a})^2-1)\right)},
\end{equation}
where $p_w$ and $\rho_w$ are the pressure and density at the wall of the rotor, $\gamma=1.067$ - adiabatic index of the working gas and $C=\sqrt{\gamma RT/\mu}$ is the sound velocity, while $\mu=0.352~ \rm kg/mol$ is the  molecular weight of $UF_6$. 
The output of the gas occurs at the bottom of the rotor (outlet). To simplify the hydrodynamics we assume that the product $P$ and waste $W$ fluxes are 
separated by a concentric  tube with radius $r_*$, which provides the specified ratio of the product flux to the feed flux $\theta=P/F$. In this case the flow lines are the straight  lines parallel to the axis. Here we neglect second order effects affecting the velocity of the gas  due to the viscous stresses and heat conduction. Therefore, $v_z$ is constant everywhere in the GC.

 $F$ is connected with the parameters of the gas as follows
\begin{equation}
 F=\int_0^a \rho v_z 2\pi rdr= {2\pi\rho_w v_z a^2 C^2\over \gamma V^2}(1-\exp({-{\gamma V^2\over 2C^2}})).
\label{eq10}
\end{equation}
The product flux $P$ is defined as 
\begin{equation}
 P=\int^{r_*}_0 \rho v_z 2\pi rdr={2\pi\rho_w v_z a^2 C^2\over \gamma V^2}(\exp({{\gamma V^2\over 2C^2}(({r_*\over a})^2-1)})-\exp({-{\gamma V^2\over 2C^2}})).
\end{equation}
Hereafter we neglect $\exp({-{\gamma V^2\over 2C^2}})$ because its value is close  $10^{-11}-10^{-14}$ for typical parameters of GC. Then,
radius $r_*$  is defined by the equation
\begin{equation}
 \theta = \exp{({\gamma V^2\over 2c^2}(({r_*\over a})^2-1))}.
\label{eq12}
\end{equation}
%

\subsection{Separative power}

Concentration $c$ in the feed flux equals to natural $c_0=7.114\cdot 10^{-3}$. We assume that $c$ does not depend on $r$ in the feed flux. 
$c$ increases near the rotational axis and reduces at the wall upon motion of the gas along the rotor. The separative power of the GC is defined as follows
\begin{equation}
 \delta U = PG(c_P)+WG(c_W)-FG(c_F),
\end{equation}
  where $G(c)=(2c-1)\ln{c\over (1-c)}$ is the separative potential introduces by Dirac and Peierls \cite{cohen}. The generalization of this equation on the case of the nonuniform flow through the feed inlet, product  and waste  outlets gives \cite{borman}
\begin{equation}
 \delta U = \int  G(c) \rho {\bf v}{d\bf S}.
\end{equation}
In our case the integration over the surface $S$ which covers all the boundaries of the working volume gives
\begin{equation}
 \delta U=2\pi \int_0^a\rho v_z(G(c)-G(c_0))rdr,
\label{eq13}
\end{equation}
where $G(c)$ is taken at the outlet. 
  
\section{Solution of the problem}

\subsection{Basic equations and assumptions}

The equation defining diffusion of the mixture of uranium isotopes in gaseous $UF_6$
is as follows
\begin{equation}
 {\partial \rho c\over \partial t}+{\partial J_k\over \partial x_k}=0,
\label{eq14}
\end{equation}
where the components $J_k$ of the flux  of $UF_6$  with the  light uranium isotope  in the cylindrical system of coordinates are given by the equations \cite{lamm}
\begin{equation}
 J_z=\rho v_z c-\rho D {\partial c\over \partial z}
\end{equation}
and 
\begin{equation}
 J_r=\rho v_r c-\rho D({\partial c\over \partial r}+{\Delta M\over M}c(1-c){\partial \ln p\over \partial r}),
\end{equation}
where $v_r$ is the radial component of the gas velocity, $c$ is the concentration of $UF_6$ with light isotope $^{235}U$.  In the case under the consideration $v_r=0$, $p$ should be taken from eq. (\ref{eq6}).
The natural concentration $c_0$ satisfies to the condition $c_0 \ll 1$. Concentration does not change strongly in the concurrent GC. Therefore  the condition $c \ll 1$ is valid everywhere and we can consider the term $(1-c)$ as a constant equal to $(1-c_0)$. Product $\rho D$ does not depend on pressure. Therefore we consider it as a constant.

Substitution of the components of $\bf J$ into (\ref{eq14}) gives the following equation for the steady state diffusion 
\begin{equation}
 {\partial \rho v_z c\over \partial z}-\rho D  {\partial^2  c\over \partial z^2}-\rho D {\partial \over r\partial r}({r\partial c\over \partial r}+\delta {\gamma V^2\over C^2}{r^2\over a^2}c)=0,
\end{equation}
where $\delta={\Delta M\over M}(1-c_0)$.

Let us introduce new variables $t=\exp{\left({\gamma V^2\over 4C^2}(({r\over a})^2-1)\right)}$,  $\tilde z=z/a$  and express concentration $c$ as 
\begin{equation}
 c=c_0 t^{-\delta}Y(t,\tilde z).
\end{equation}
In these variables density $\rho =\rho_w t^2$. 
For $Y$ we obtain the following equation in these variables
\begin{equation}
 t{\partial\over \partial t}t{\partial Y\over \partial t}-4t^2 {\rho_w v_z a\over \rho D}({C^2\over \gamma V^2})^2 {\partial Y\over \partial \tilde z}+4({C^2\over \gamma V^2})^2{\partial^2 Y\over \partial \tilde z^2}-\delta^2 Y=0.
\label{eq19}
\end{equation}
Here we neglect variation of $ (r/a)^2$. This is reasonable because the geometrical scale $\Delta r$ on which  density, pressure and $t$ vary along radius is small compared with $a$. Indeed,
\begin{equation}
 {\Delta r\over a}={C^2\over \gamma V^2}. 
\end{equation}
For $UF^6$ typical $C=86~ \rm m/s$, while $V > 600~ \rm m/s$. This means that  ${\Delta r\over a} < 2\cdot 10^{-2}$.   

Let us express $\rho_w$ in (\ref{eq19}) trough $F$ using (\ref{eq8}). After that eq.  (\ref{eq19}) takes a form
\begin{equation}
 t{\partial\over \partial t}t{\partial Y\over \partial t}-t^2 {2F\over\pi a \rho D}({C^2\over \gamma V^2}) {\partial Y\over \partial \tilde z}+4({C^2\over\gamma V^2})^2{\partial^2 Y\over \partial \tilde z^2}-\delta^2 Y=0.
\label{eq21}
\end{equation}
Here it is convenient to replace variable $\tilde z$ on variable $\xi$  according to the equation
\begin{equation}
 \xi= {\gamma V^2\over 2C^2}\tilde z.
\label{eq22}
\end{equation}
Then we obtain
\begin{equation}
 t{\partial\over \partial t}t{\partial Y\over \partial t}-t^2 {F\over\pi a \rho D}{\partial Y\over \partial \xi}+{\partial^2 Y\over \partial \xi^2}-\delta^2 Y=0.
\label{eq23}
\end{equation}
This equation can be simplified. Distribution of $Y$ is as follows at the inlet
\begin{equation}
Y(t,0)=t^{\delta}.
\label{eq24}
\end{equation}
 Maximal  possible variation of $Y$ corresponds to an infinitely long rotor, where the concentration of light isotope  goes to the nonuniform distribution defined by equation 
\begin{equation}
 Y(t,\infty)=t^{-\delta}.
\end{equation}
The main contribution into separative work gives the concentration at  radius $r_*$ separating the product and waste fluxes because the mass flux density in the product flux is maximal here. The variation $\Delta Y$ in this case is as follows
\begin{equation}
 \Delta Y= \theta ^{-\delta}- \theta ^{\delta},
\end{equation}
where $\theta = t_*$ is the value of $t_*$ at the $r_*$.  Taking into account that $\theta ^{\delta} \approx 1$ we have an estimate
\begin{equation}
 \Delta  Y \approx 2\delta \ln({1\over \theta}).
\end{equation}

Now let us take into account that  $ {\partial^2 Y\over \partial \xi^2} \approx {\Delta Y\over \xi^2_{L}}$, where $ \xi_{L}={2C^2\over \gamma V^2}{L\over a}$. The term $ {\partial^2 Y\over \partial \xi^2} $  can be neglected in eq. (\ref{eq23}) provided that it is much less the last term $\delta^2 Y$ in this equation. This condition gives
\begin{equation}
 2({2C^2\over \gamma V^2}{L\over a})^2\ln({1\over \theta}) \ll \delta. 
\label{eq28}
\end{equation}
The last condition  is fulfilled for reasonable parameters of the centrifuges.   This means that in the concurrent centrifuge we can neglect the axial diffusion in compare with the radial one provided that condition (\ref{eq28}) takes place.   The basic equation for the concentration takes the following form in this case 
\begin{equation}
 t{\partial\over \partial t}t{\partial Y\over \partial t}-t^2 {F\over\pi a \rho D}{\partial Y\over \partial \xi}-\delta^2 Y=0.
\label{eq28a}
\end{equation}

\subsection{Boundary conditions}

Boundary condition at the inlet  (feed flux) follows to distribution (\ref{eq24}). At the outlet end of the rotor no boundary conditions should be specified because we neglect the axial diffusion and the concentration is advected to this boundary from the computational domain.  Boundary condition at the wall of the rotor corresponds to zero flux of the concentration 
\begin{equation}
 {\partial c\over \partial r}+\delta c{\partial \ln p\over \partial r}=0.
\label{eq29}
\end{equation}
Substitution of the pressure distribution (\ref{eq6})  into this equation gives  the following boundary condition for function $Y$
\begin{equation}
 {\gamma V^2\over 2C^2}{r\over a^2}  t^{-\delta}(t{\partial Y\over \partial t}+\delta Y)=0.
\end{equation}
At the wall of the rotor ($t=1$) this gives the following boundary condition 
\begin{equation}
 t{\partial Y\over \partial t}+\delta Y=0.
\label{eq31}
\end{equation}

We formally extend  the variation of $t$ from 1 at the wall of the rotor to 0, because at the axis  $t$ is of the order $10^{-11} - 10^{-14}$. We also neglect the variation of $r/a$ assuming that this  parameter equals to 1. Therefore, the boundary condition at $t=0$ should be 
\begin{equation}
 t^{-\delta}(t{\partial Y\over \partial t}+\delta Y)=0.
\label{eq32}
\end{equation}

\subsection{Solution}

We can expand the solution of eq. (\ref{eq28a}) on functions $\exp(-k\xi)$ in a sum of series   as follows
 \begin{equation}
  Y(t,\xi)=\sum \exp(-k_i\xi)S_i(\xi), 
\label{eq33}
 \end{equation}
where functions $S_i$ satisfy to the equation
\begin{equation}
 {\partial\over t\partial t}t{\partial S_i\over \partial t}+ ({F k_i\over\pi a \rho D}-{\delta^2\over t^2}) S_i=0.
\label{eq34}
\end{equation}
General solution of this equation can be expressed as a linear combination of Bessel functions $J_{\delta}(\lambda t)$ and $J_{-\delta}(\lambda_i t)$
\begin{equation}
 S_i=E J_{\delta}(\lambda_i t)+B J_{-\delta}(\lambda_i t),
\end{equation}
where $E$ and $B$ are some constants and $\lambda_i= \sqrt{{F k_i\over\pi a \rho D}}$.  To define these constants let us firstly consider the boundary condition at $t=0$. 
According to \cite{abram} the functions $J_{\delta}(\lambda_i t)$ and $J_{-\delta}(\lambda_i t)$ behave near the point $t=0$ as 
\begin{equation}
 J_{\delta}(\lambda_i t)= {1\over \Gamma(\delta+1)}({\lambda_i t\over 2})^{\delta}, ~~J_{-\delta}(\lambda_i t)= {1\over \Gamma(1-\delta)}({\lambda_i t\over 2})^{-\delta}(1-{(\lambda_i t)^2\Gamma(1-\delta)\over 4\Gamma(2-\delta)}).
\end{equation}
Thus, one of the functions is regular in this point, while another diverges. Substitution of the function $J_{\delta}$ into boundary condition (\ref{eq32}) gives the expression
\begin{equation}
{2\delta\over  \Gamma(1+\delta)}({\lambda_i\over 2})^2
\end{equation}
which does not equal to $0$.  Substitution of $J_{-\delta}$ into (\ref{eq32}) gives the expression 
\begin{equation}
 -{2\lambda^{2-\delta} t^{2-2\delta}\over 2^{2-\delta}\Gamma(2-\delta)},
\end{equation}
which goes to $0$ at $t \rightarrow 0$. Thus, $S_i=B J_{-\delta}(\lambda_i t)$.
Boundary condition at the wall of the rotor gives
\begin{equation}
 \lambda_i{\partial J_{-\delta}(\lambda_i)\over \partial \lambda_i}+\delta J_{-\delta}(\lambda_i)=0.
\end{equation}
It follows from the properties of Bessel functions that (\cite{abram}) 
\begin{equation}
 t{\partial J_{-\delta}(t)\over \partial t}+\delta J_{-\delta}(t)=-J_{1-\delta}(t).
\end{equation}
 Therefore, the boundary condition at the rotor wall gives us the equation defining the eigenvalue of the problem 
\begin{equation}
 J_{1-\delta}(\lambda_i) = 0.
\end{equation}
Four first  eigenvalues are $\lambda_0= 0;~\lambda_1=3.82;~\lambda_2=7;~\lambda_3=10.15$. 

The eigenfunction at $\lambda=0$ can be defined directly from eq. (\ref{eq34}) as follows
\begin{equation}
 S_0(t)=B_0 t^{-\delta}.
\end{equation}

Finally the solution takes the form
\begin{equation}
 Y(t,\xi)=B_0 t^{-\delta}+ \sum_{i=1}^{i=\infty} B_i \exp{(-k_i\xi)}J_{-\delta}(\lambda_i t).
\end{equation}
Coefficients $B_i$ in the expansion of the solution into the sum of the series (\ref{eq33}) are defined from the boundary condition at the inlet $\xi=0$. This gives
\begin{equation}
 t^{\delta}=B_0 t^{-\delta}+ \sum_{i=1}^{i=\infty} B_i J_{-\delta}(\lambda_i t).
\end{equation}
Eigenfunctions $J_{-\delta}(\lambda_i t)$ are orthogonal  to each other and to $t^{-\delta}$ with weight function $t$. Therefore, multiplication of this equation on $t^{1-\delta}$ and integration over $t$ gives the following value of $B_0$ 
\begin{equation}
 B_0=1-\delta.
\end{equation}
Coefficients $B_i$ are defined from the equations
\begin{equation}
 B_i={\int_0^1 t^{\delta}J_{-\delta}(\lambda_i t)tdt\over \int_0^1 J^2_{-\delta}(\lambda_i t)tdt}.
\end{equation}
It can be shown that for the specified  boundary conditions at $t=0$ and at $t=1$ we have 
\begin{equation}
 \int_0^1 t^{\delta}J_{-\delta}(\lambda_i t)tdt={1\over \lambda^{2+\delta}_i}\left({2^{1+\delta}\over \Gamma(-\delta)}-\lambda_i^{1+\delta}J_{-(1+\delta)}(\lambda_i)\right),
\end{equation}
and 
\begin{equation}
 \int_0^1 J^2_{-\delta}(\lambda_i t)tdt={1\over 2}J^2_{-\delta}(\lambda_i).
\end{equation}

\section{Optimized separation power}

\begin{figure} 
\begin{center}
\includegraphics[width=120mm]{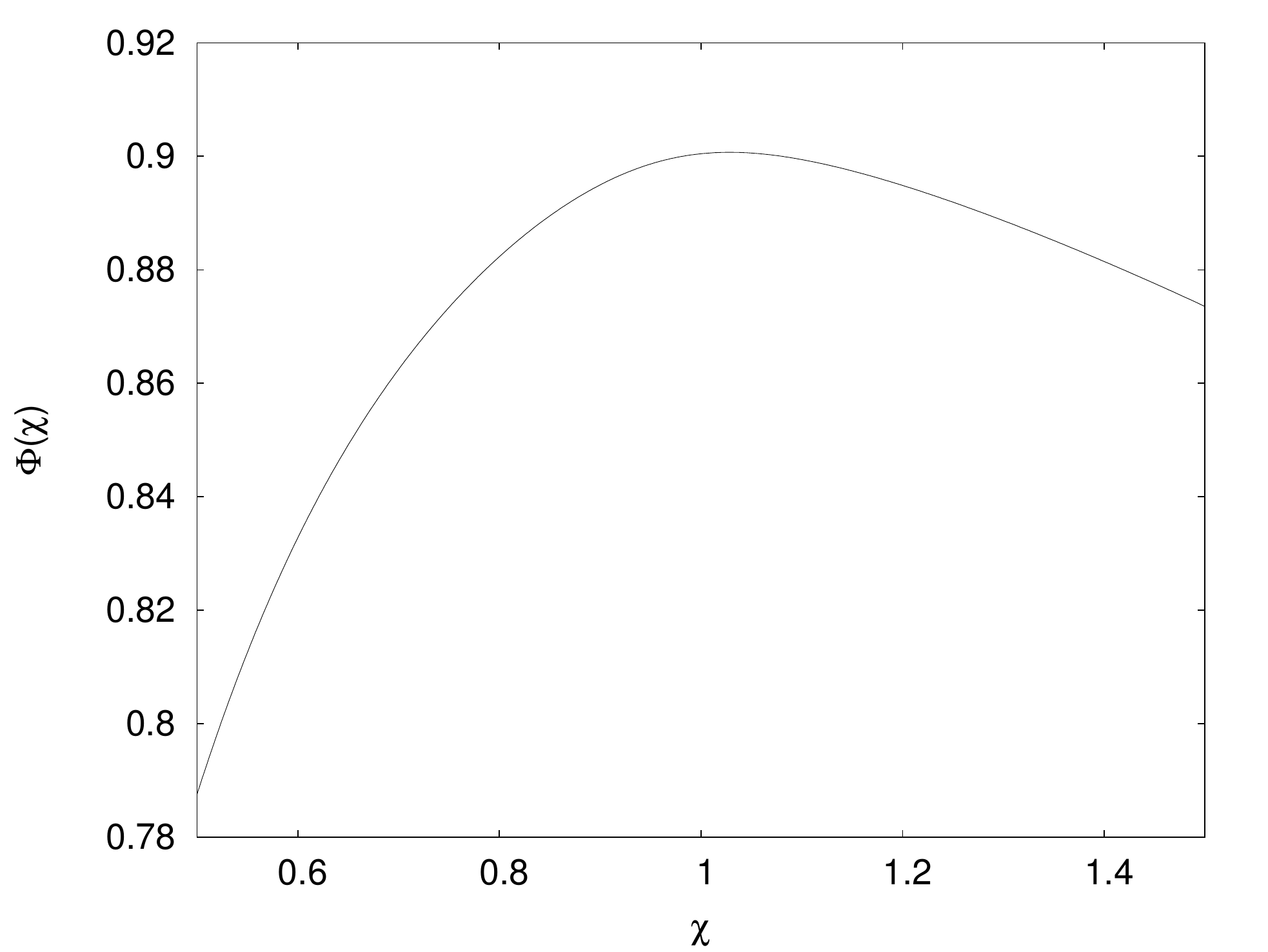}
\caption{Dependence of $\Phi$ on $\chi$ at $c_0=7.114\cdot 10^{-3}$ and $\delta={3\over 352}(1-c_0)$}
 \label{f2}
\end{center}
\end{figure}

It follows from  eqs. (\ref{eq8}) and (\ref{eq10}) that 
\begin{equation}
2\pi \rho v_z rdr= 2Ftdt. 
\end{equation}
Therefore the separative power (\ref{eq13}) can be presented as 
\begin{equation}
 \delta U= F\int_0^1(G(c)-G(c_0))2tdt, 
\label{eq51}
\end{equation}
where integration is performed over outlet. It is convenient to express $k_i$ as 
\begin{equation}
 k_i={\pi a \rho D\over F}\lambda_i^2.
\end{equation}
In this case the solution can be presented as 
\begin{equation}
 c(t,\xi)=(1-\delta)t^{-2\delta}\left(1+\sum_1^{\infty}{B_i\over (1-\delta)}t^{\delta}\exp{ (-{\pi a \rho D\over F}\lambda_i^2\xi)}J_{-\delta}(\lambda_i t)\right). 
\label{eq53}
\end{equation}
Let us introduce new  variable $\chi$ as follows
\begin{equation}
 F=\pi a \rho D \lambda_1^2\xi \chi.
\label{eq56}
\end{equation}
Then the solution takes a form 
\begin{equation}
 c(t,\chi)=(1-\delta)t^{-2\delta}\left(1+\sum_1^{\infty}{B_i\over (1-\delta)}t^{\delta}\exp{ (-{\lambda_i^2\over \lambda_1^2\chi})}J_{-\delta}(\lambda_i t)\right),
\end{equation}
and the separative power becomes
\begin{equation}
  \delta U= \pi a \rho D \lambda_1^2\xi \chi\int_0^1(G(c)-G(c_0))2tdt.
\end{equation}
Substitution of (\ref{eq22}) into this  equation at $z=L$ gives the following expression for $\delta U$
\begin{equation}
  \delta U= \pi \left({\delta\over 2}\right)^2 \rho D \lambda_1^2 {\gamma V^2\over 2 C^2}L\Phi(\delta,c_0,\chi),
\label{eq57}
\end{equation}
where function $\Phi$ is
\begin{equation}
 \Phi(\delta,c_0,\chi)=\chi\left({2\over \delta}\right)^2 \int_0^1(G(c)-G(c_0))2tdt.
\end{equation}
This function depends only on 3 parameters $\delta, ~c0$ and $ \chi$. {\bf Dependence of $\Phi$ on $\delta$ and $c0$ can be neglected  at $\delta \ll 1$ and $c0 \ll 1$. These parameters  are specified  by the properties of the working gas}.  The only variable parameter is $\chi$. This is the dimensionless feed flux. Optimization of the concurrent centrifuge is possible only on this flux. This has clear physical sense. At the fixed length of the centrifuge $L$ and pressure a small feed fluxes result into a small velocity $v_z$. Slow advection results into maximal possible separation of the isotopes. In this conditions decrease of the feed flux results
into decrease of $\delta U$ because of the first term in the right hand side of eq. (\ref{eq51}). In the opposite case of large feed flux, velocity $v_z$ can be so large that the separation is negligible in compare with the advection of the concentration to the outlet. Further increase of the feed flux results only to the decrease of the separation. At very large feed fluxes we also have to expect decrease of $\delta U$ because of decrease of the integral in the right hand side of   eq.   (\ref{eq51}). Therefore, $\delta U$ have to have a maximum at some feed flux which correspond  to the optimal feed flux in the centrifuge.       
The dependence of $\Phi$ on $\chi$ is shown in fig. \ref{f2}. This function has a maximum equal to $\sim 0.9 $ at $\chi = 1$.

To express the optimized separative power of the centrifuges  in conventional Separation Work Units (SWU) per year it is necessary to multiply eq. (\ref{eq57}) on 1 year in seconds and on a factor $238\over 352$ of ratio of weight of metallic uranium over weight  of the working gas. After that  
we obtain 
\begin{equation}
 \delta U = 12.7 \left({V\over 700~ \rm m/s}\right)^2 \left({300~{\rm K}\over T}\right) L , ~\rm kg\cdot  SWU/year
\label{eq59}
\end{equation}
At the calculation of the separative power we assumed conventional value for $\rho D=2.3\cdot 10^{-5}~\rm Pa\cdot s$.

\section{Discussion of the results}

{\bf In this work we have solved analytically the problem of diffusion of the uranium isotope mixture in the prescribed uniform flow of the gas with an exponential profile of density along radius in the concurrent centrifuge.  The separative power of the GC has been calculated and optimized on the basis of this solution. For the first time an analytical expression for the optimized separative power of the gas centrifuge is obtained practically from the first principles which agree with experimental data (compare eqs. (\ref{eq3}) and (\ref{eq59})) }.
Equation (\ref{eq59}) shows that in contrast to the result by Cohen \cite{cohen}  the optimized separative power of the concurrent gas centrifuge is proportional to $V^2 L$. This dependence well reproduces the experimental data for Russian counter current centrifuges approximated by eq. (\ref{eq3}).
Surprisingly, the numerical coefficient  in this dependence coincides with the experimental one in the limits of uncertanties. The separative power of the concurrent centrifuge does not depend on the rotor diameter. This is the only difference between equations (\ref{eq3}) and (\ref{eq59}). 

It is interesting to understand why the optimized separative power of the concurrent centrifuge depends on the velocity of the rotor rotation as $V^2$ and this does not contradict to eq. (\ref{eq4}). It follows from eq. (\ref{eq53}) that maximal value of $q-1$ equals to
\begin{equation}
 q-1=t_{*}^{-2\delta}-1=2\delta\ln{1\over \theta},
\label{eq60}
\end{equation}
where $t_*=\theta$ (see eq. (\ref{eq12})).
If we kept the separation of the fluxes $\theta$ constant upon increase of $V$ the value $q-1$ remains constant as well. According to eq. (\ref{eq12})  this means that
upon growth of $V$ the radius $r_*$ of the concentric tube separating the fluxes  increases, keeping the product 
\begin{equation}
 \left({V\over C}\right)^2\left(({r_*\over a})^2-1\right)
\end{equation}
constant.  As the result all the growth of $\delta U$ with $V$ is due to growth of the feed flux. This flux increases with $V$ because the rate of the radial separation increases as $V^2$. Therefore, it is possible to increase the mass flux proportional to $V^2$ to provide efficient separation.  The same is valid for dependence of $\delta U$ on $L$. $q$ and $\theta$  do not depend on $L$. The feed flux increases proportionally to $L$ because the working gas has  more traveling time  for radial separation.   

The optimized  separative power does not depend on the pressure of the gas at the wall of the rotor. This is another important feature of the obtained solution. It is clear why this occurs.  Everywhere  density enters  into the  equations in combination $\rho v_z$. Product $\rho D$ depends only on temperature, not density. Correspondingly,  solution (\ref{eq53}) depends on the feed flux $F$. Due to eq. (\ref{eq10}) the optimal separative power is  not unique on the plane of parameters $p_w,v_z$. Combining eqs. (\ref{eq10}) and  (\ref{eq56}) at $\chi=1$ we obtain that the optimal $\delta U$ is constant along the  line located on the plane  $p_w, ~v_z$ defined by the equation
\begin{equation}
 p_w \cdot v_z={1\over 4}{\rho D \lambda_1^2 L V^2\over a^2} \left({\gamma V^2\over  C^2}\right).
\end{equation}  
 In the case of specification of pressure the axial velocity will be
\begin{equation}
 v_z=60
\left({100~ \rm mm ~Hg\over p_w}\right) \left({6~ \rm cm\over a}\right)^2\left({V \over 700~ \rm m/s}\right)^4\left({300~ \rm K\over T}\right)L,~ {\rm m/s}.  
\end{equation}

Optimal feed flux equals to 
\begin{equation}
 F=\pi  \rho D \lambda_1^2 {\gamma V^2\over 2 C^2}L=35 \left({V\over 700~ \rm m/s}\right)^2 L,~ ~\rm g/s. 
\end{equation}
This means  that  the optimal concurrent centrifuge needs very large feed flux and provides very  small coefficient of separation defined by eq. 
(\ref{eq60}). The electric power necessary to provide rotation of this feed flux can be estimated as  
\begin{equation}
 W=F {V^2\over 2}=8.6 \left({V\over 700~ \rm m/s}\right)^4 {300 {\rm K}\over T}L,  ~\rm kWatt.
\end{equation}
Actually, this is the minimal necessary electric power  spent for the work of the centrifuge. If to compare with the concurrent centrifuge, the production of the same separative work the concurrent centrifuge demands much more energy. This is the reason why the concurrent centrifuges are not explored for the industrial production of uranium isotopes.

\section{Acknowledgements}

The work has been performed under support of  Ministry of education and science of Russia, grant no. 3.726.2014/K





\end{document}